 \documentclass[final,authoryear,3p,times]{elsarticle}

\usepackage{amssymb,amsmath,graphicx,setspace,multirow,color,rotating,subfigure,ulem,url}
\usepackage{mathrsfs}
\usepackage{textcomp}
\usepackage{float}
\usepackage{booktabs}

\journal{XX}

\begin{document}

\begin{frontmatter}

\title{Modeling aggressive market order placements with Hawkes factor models
}
\author[BS]{Hai-Chuan Xu}
\author[BS,SS]{Wei-Xing Zhou \corref{cor}}
\cortext[cor]{Corresponding author.}
\ead{wxzhou@ecust.edu.cn} %

\address[BS]{School of Business and Research Center for Econophysics, East Chine University of Science and Technology, Shanghai 200237, China}
\address[SS]{Department of Mathematics, East China University of Science and Technology, Shanghai 200237, China}

\begin{abstract}
Price changes are induced by aggressive market orders in stock market. We introduce a bivariate marked Hawkes process to model aggressive market order arrivals at the microstructural level. The order arrival intensity is marked by an exogenous part and two endogenous processes reflecting the self-excitation and cross-excitation respectively. We calibrate the model for an SSE stock. We find that the exponential kernel with a smooth cut-off (i.e. the subtraction of two exponentials) produces much better calibration than the monotonous exponential kernel (i.e. the sum of two exponentials). The exogenous baseline intensity explains the $U$-shaped intraday pattern. Our empirical results show that the endogenous submission clustering is mainly caused by self-excitation rather than cross-excitation.
\end{abstract}


\end{frontmatter}


\section{Introduction}

Self-exciting and mutually exciting point processes are a natural extension of Poisson processes, which are first proposed by Alan G. Hawkes \citep{Hawkes-1971-JRSSB,Hawkes-1971-Bm}. Hawkes processes have been applied to characterize clustering events in finance, particularly to high-frequency data and market microstructure \citep{Bacry-Mastromatteo-Muzy-2015-MML,Hawkes-2018-QF}, because many types of events are clustered in time such as order submissions \citep{Large-2007-JFinM}, mid-quotes changes \citep{Filimonov-Sornette-2012-PRE}, transactions \citep{Lallouache-Challet-2016-QF}, and extreme returns occurrences \citep{Bormetti-Calcagnile-Treccani-Corsi-Marmi-Lillo-2015-QF}.

As a class of branching processes, self-exciting Hawkes models can be used to compute the so-called branching ratio, which is defined as the average number of triggered events of the first generation per source \citep{Hardiman-Bercot-Bouchaud-2013-EPJB,Saichev-Sornette-2014-PRE,Filimonov-Sornette-2015-QF}. \cite{Chavez-Demoulin-Mcgill-2012-JBF} propose a marked self-exciting process model to feature intraday clustering of extreme fluctuations and calculate the instantaneous conditional VaR. \cite{Filimonov-Bicchetti-Maystre-Sornette-2014-JIMF} calibrate the self-exciting Hawkes model on time series of price changes and then quantify the endogeneity and structural regime shifts in commodity markets. Similarly, \cite{Filimonov-Wheatley-Sornette-2015-CNSNS} propose an effective measure of endogeneity for the autoregressive conditional duration point processes. \cite{Blanc-Donier-Bouchaud-2017-QF} introduce Quadratic Hawkes models by allowing all feedback effects in the jump intensity that are linear and quadratic in past returns.

In addition to self-exciting processes, more researchers study the cross-exciting effects through multivariate Hawkes processes. \cite{Bowsher-2007-JEm} proposes a bivariate point process model of the timing of trades and mid-quote changes for a New York Stock Exchange stock. \cite{Large-2007-JFinM} uses a ten-variate Hawkes process to measure the resilience of London Stock Exchange order book. \cite{Muni-Pomponio-2012-EEJ} show that a simple bivariate Hawkes process fits nicely their empirical observations of trades-through. \cite{Bacry-Muzy-2014-QF} introduce a multivariate Hawkes process with 4 kernels that accounts for the dynamics of market prices. \cite{Zheng-Roueff-Abergel-2014-SIAMjfm} also introduce a multivariate point process describing the dynamics of the bid and ask price of a financial asset. \cite{Rambaldi-Pennesi-Lillo-2015-PRE} present a Hawkes process to model high-frequency price dynamics in the foreign exchange market. The price change is affected by a self-exciting mechanism and an exogenous component generated by the pre-announced arrival of macroeconomic news. \cite{Ait-Sahalia-Cacho-Diaz-Laeven-2015-JFE} model financial contagion across six international equity index using mutually exciting jump processes. \cite{Bacry-2016-QF} presents a non-parametric Hawkes kernel estimation procedure and apply it to high-frequency order book modelling. \cite{Rambaldi-Bacry-Lillo-2017-QF} propose multivariate Hawkes processes to study complex interactions between the time of arrival of orders and their size. \cite{Calcagnile-Bormetti-Treccani-Marmi-Lillo-2018-QF} explain the synchronization of large price movements across assets using a multivariate point process. Furthermore, \cite{Lu-Abergel-2018-QF} extend nonlinear Hawkes processes to describe limit order books.

In this paper, we are interested in modelling aggressive market order placement, i.e. orders with the size greater than the opposite best quote. These orders consume liquidity and walk up the limit order book, causing the best-quotes to change. Aggressive market orders are very important in price formation and microstructure. For example, the submission pattern of aggressive market orders may contain information about order splitting behavior according to the liquidity available in the order book. We introduce a bivariate marked Hawkes process to model aggressive market order arrivals. It's reasonable to apply a Hawkes process to model the order events because the inter-trade durations have fat tails and long memory \citep{Jiang-Chen-Zhou-2008-PA,Jiang-Chen-Zhou-2009-PA,Ruan-Zhou-2011-PA}. We find that the exponential kernel with a smooth cut-off (i.e. the subtraction of two exponentials) at short times  produces better calibration than the monotonous exponential kernel (i.e. the sum of two exponentials) does. Our empirical results show that the endogenous submission clustering is mainly caused by self-excitation rather than cross-excitation. In addition, the exogenous aggressive order arrivals show obvious intraday pattern.

The paper is organized as follows. Section \ref{Sec:Model} describes the bivariate marked Hawkes process for aggressive market order placements, including kernel functions, stationarity condition, parametric estimation and mis-specification testing. Section \ref{Sec:Data} describes the empirical data we use and analyzes the inter-arrival durations. Section \ref{Sec:Calibration} calibrates the Hawkes models and shows empirical results. Section \ref{Sec:Conclusion} summarizes.

\section{Hawkes model}
\label{Sec:Model}

\subsection{Bivariate marked Hawkes process}

Let $N_1$ and $N_2$ denote the counting processes for aggressive market buy orders and aggressive market sell orders. These two processes are assumed to form a bivariate Hawkes process with intensities $\lambda_1$ and $\lambda_2$,
\begin{equation}\label{Eq:Model}
  \begin{array}{ccl}
  \lambda_1(t) &=& \mu_1(t) + \int_0^t\phi_{11}(t-s)dN_1(s) + \int_0^t\phi_{12}(t-s)dN_2(s),\\
  \\
  \lambda_2(t) &=& \mu_2(t) + \int_0^t\phi_{21}(t-s)dN_1(s) + \int_0^t\phi_{22}(t-s)dN_2(s),
  \end{array}
\end{equation}
where $\mu_i>0$ is a baseline intensity describing the arrival of exogenous events, and the kernels $\phi_{ii}$ and $\phi_{ij}$ represent respectively the self-exciting and cross-exciting effects.

\begin{figure}
  \centering
  \includegraphics[width=0.5\linewidth]{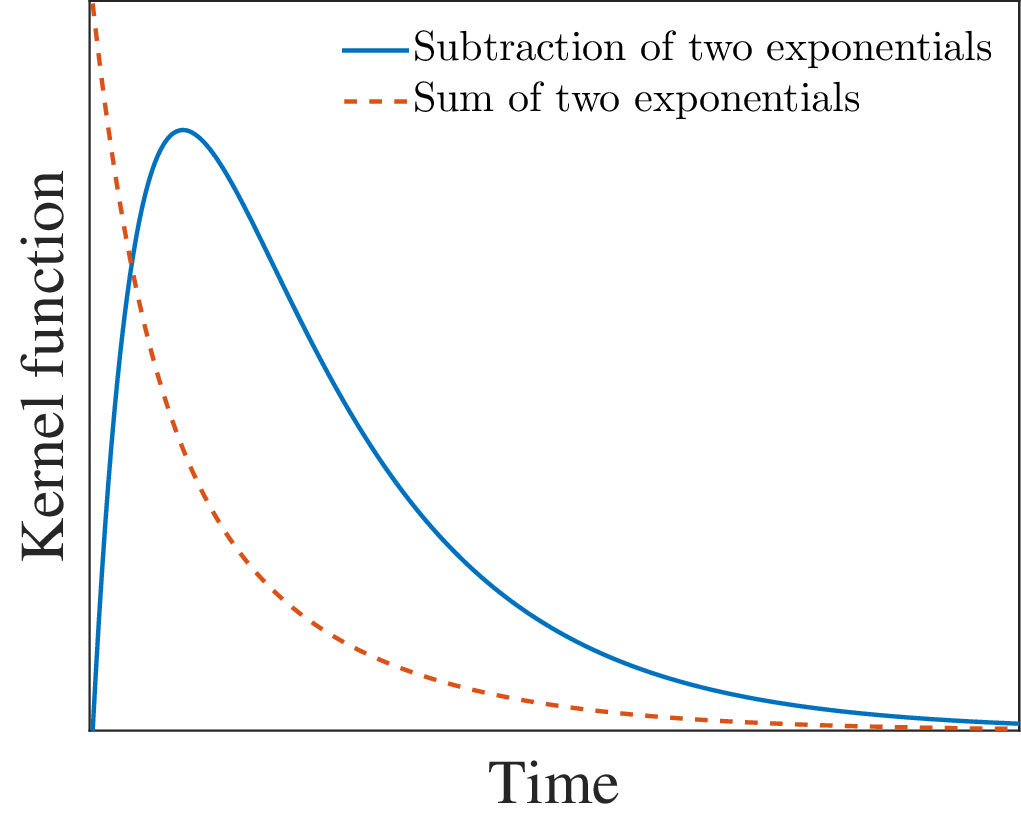}
  \caption{Illustration of the kernel with a subtraction of two exponentials (continuous blue line) and the kernel with a sum of two exponentials.}
  \label{Fig:illustration:kernel}
\end{figure}

The kernel $\phi(t-s)$ describes the impact of a previous event at time $s$ on the current intensity at time $t$. Previous studies advocate the use of exponential or power-law kernels. Here we use the difference of two exponentials as the kernel to account for the self- or cross-excitations:
\begin{equation}\label{Eq:biexponential:kernel}
  \phi_{ij}(u)=k_{ij}e^{b_{ij}v_j}[e^{-\alpha_{ij}u} - e^{-\beta_{ij}u}],
\end{equation}
where $v_j$ is the share volume of the order event. The negative exponential term provides a smooth cut-off at short times. This function has several advantages. First, it satisfies $\phi_{ij}(0)=0$, since we cannot expect market participants to react instantaneously to events. Second, it allows excitations smoothly increase to the highest and then gradually fade over time (see Fig.~\ref{Fig:illustration:kernel}), which is more reasonable to characterize the reaction of market participants. \cite{Hardiman-Bercot-Bouchaud-2013-EPJB} and \cite{Filimonov-Sornette-2015-QF} also suggest similar kernel functions. In our empirical analysis, we will compare the kernel in Eq.~(\ref{Eq:biexponential:kernel}) with the sum of exponentials kernel below
\begin{equation}\label{Eq:biexponential:kernel:sum}
  \phi_{ij}(u)=k_{ij}e^{b_{ij}v_j}[e^{-\alpha_{ij}u} + e^{-\beta_{ij}u}].
\end{equation}
In this case, we also set $\phi_{ij}(0)=0$ like other literatures. Third, compared with power-law kernels, the use of exponential kernels can reduce the computational complexity from $\mathcal{O}(N^2)$ to $\mathcal{O}(N)$.

\subsection{Stationarity condition}

A multivariate point process is stationary if the joint distribution of any number of types of events on any number of given intervals is invariant under translation. According to Theorem 7 in \cite{Bremaud-Massoulie-1996-AP}, the stationarity condition of a multivariate point process is that, the matrix $Q$ with entries $q_{ij}=\int_0^{+\infty}|\phi_{ij}(u)|du$ has a spectral radius strictly less than 1. For our bivariate Hawkes model with the kernel in Eq.~(\ref{Eq:biexponential:kernel}), the matrix $Q$ is
\begin{equation}\label{Eq:stationarity:matrix}
  Q = \left[
  \begin{array}{ll}
  k_{11}e^{b_{11}v_1}(\frac{1}{\alpha_{11}}-\frac{1}{\beta_{11}})~~~&  k_{12}e^{b_{12}v_2}(\frac{1}{\alpha_{12}}-\frac{1}{\beta_{12}})\\
  \\
  k_{21}e^{b_{21}v_1}(\frac{1}{\alpha_{21}}-\frac{1}{\beta_{21}})~~~&  k_{22}e^{b_{22}v_2}(\frac{1}{\alpha_{22}}-\frac{1}{\beta_{22}})
  \end{array}
  \right].
\end{equation}
In the same way, the matrix $Q$ for the kernel in Eq.~(\ref{Eq:biexponential:kernel:sum}) is
\begin{equation}\label{Eq:stationarity:matrix:sum}
  Q = \left[
  \begin{array}{ll}
  k_{11}e^{b_{11}v_1}(\frac{1}{\alpha_{11}}+\frac{1}{\beta_{11}})~~~&  k_{12}e^{b_{12}v_2}(\frac{1}{\alpha_{12}}+\frac{1}{\beta_{12}})\\
  \\
  k_{21}e^{b_{21}v_1}(\frac{1}{\alpha_{21}}+\frac{1}{\beta_{21}})~~~&  k_{22}e^{b_{22}v_2}(\frac{1}{\alpha_{22}}+\frac{1}{\beta_{22}})
  \end{array}
  \right].
\end{equation}
We recall that the spectral radius of the matrix $Q$ is defined as $\rho(Q)=\max_{a\in\ell(Q)}|a|$, where $\ell(Q)$ denotes the set of all eigenvalues of $Q$.

\subsection{Parameter estimation}

For each type of orders, $\mu_i$ is taken as a seasonal piecewise linear spline with 4 knots at 9:30am, 10:00am, 10:30am, 11:30am for the morning session, and 13:00pm, 14:00pm, 14:30pm, 15:00pm for the afternoon session. Therefore, the intensity $\lambda_i$ dependents on the following parameter set $\theta_i$,
\begin{equation}\label{Eq:parameters}
  \theta_i = \{k_{ij},b_{ij},\alpha_{ij},\beta_{ij}, {\mathrm{knots~of}}~\mu_i : j = 1,2\}.
\end{equation}
In other words, there are 12 parameters to be estimated for each type of orders and for each time interval. Suppose that the data is observed over the interval $[0,T]$, then the maximum likelihood estimates for $\theta_i$ can be obtained by maximizing
\begin{equation}\label{Eq:MLE}
  \int_0^T -\lambda_i(u)du + \sum_{n:t_{n,i}<T}\log \lambda_i(t_{n,i}),
\end{equation}
where $\{t_{n,i}\}$ is the sequence of the times of the events of type $i$ (see Theorem 3.1 in \cite{Bowsher-2007-JEm}).

\subsection{Mis-specification testing}

The quality of the fits is then assessed on the time-deformed series of durations $\{\tau_{n,i}\}$, defined by
\begin{equation}\label{Eq:mis:specification}
  \tau_{n,i} = \int_{t_{n-1,i}}^{t_{n,i}}\hat{\lambda_i}(u)du,
\end{equation}
where $\hat{\lambda_i}$ is the estimated intensity and $\{t_{n,i}\}$ are the empirical time stamps. If a Hawkes process describes the data correctly, the values of $\tau_{n,i}$ must be independent and exponentially distributed with the rate equal to 1. This can be verified visually in QQ-plots and rigorously with the Kolmogorov-Smirnov test \citep{Bowsher-2007-JEm}.

\section{Data description}
\label{Sec:Data}


We use order flow data of the stock China Vanke (000002.SZ) traded on the Shenzhen stock exchange from April 10th, 2003 to May 20th, 2003. We choose these 21 days data due to high activity of order events around the annual financial report announcement. In this paper we consider the order flow occurring in the continuous double auction period (9:30 AM to 11:30 AM and 1:00 PM to 3:00 PM). Note that there is a lunch effect, that is, the morning orders nearly have no impact on the submission of afternoon orders after 1.5 hours lunch break. Therefore, we will estimate our model separately for the morning and afternoon sessions.

A problem arises regarding the granularity of the data. Because time stamps are rounded to the nearest 10 milliseconds, the data set contains multiple events with the same time stamp. For our sample, there are only 0.61\% of events having the same time stamp as some other events. Comparing with the data in \cite{Large-2007-JFinM}, which are rounded to the nearest seconds, there are 40\% of events have the same time stamp. Due to ignorable probability to have multiple events in our sample, we do not handle the date and assume that each event occurring within 10 milliseconds is independent of all the others if any within the same interval.

We mark the impact of order size $v_i$ in our model, so we calculate their median values: 3000 shares for aggressive market buy orders and 3200 shares for aggressive market sell orders. We also calculate the proportions of order penetration, which is the number of price levels on the opposite order book that the order consumes. The results are presented in Table~\ref{Tb:Penetration}. We can see that most aggressive market orders consume only the orders at the first price level.

\begin{table}[h]
  \centering
  \caption{Proportions of market orders with different penetrations. The proportions of orders with the penetration greater than 9 are zeros.}
  \begin{tabular}{ccccccccc}
    \hline\hline
    Penetration & 1 & 2 & 3 & 4 & 5 & 6 & 7 & 8 \\
    \hline
    Market buys  & 86.70\% & 10.47\% & 2.24\% & 0.38\% & 0.11\% & 0.06\% & 0.04\% & 0.00\% \\
    Market sells & 84.89\% & 11.66\% & 2.54\% & 0.66\% & 0.16\% & 0.05\% & 0.02\% & 0.02\% \\
    \hline\hline
  \end{tabular}
  \label{Tb:Penetration}
\end{table}

\section{Empirical implementation}
\label{Sec:Calibration}

\begin{figure}[h]
\centering
  \includegraphics[width=0.45\linewidth]{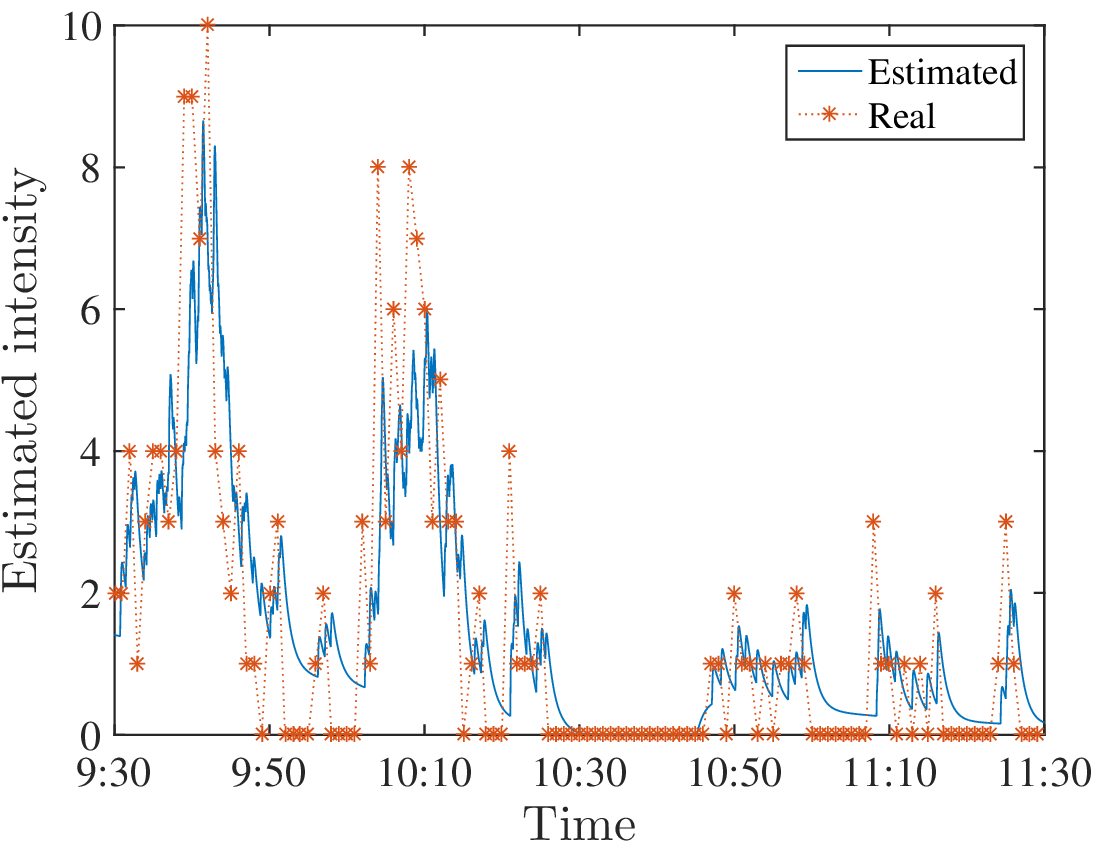}
  \caption{(color online) The estimated intensity of aggressive market buy orders in the morning of April 10th, 2003 (events per minute) (blue line). The kernel function used in Hawkes model is given in Eq.~(\ref{Eq:biexponential:kernel}). The real intensity, i.e. the number of aggressive market buy orders in every minute, is charted by orange stars.}
  \label{Fig:Estimated:Intensity}
\end{figure}

In Fig.~\ref{Fig:Estimated:Intensity}, we present a sample of the estimated intensity path of aggressive market buy orders in the morning of April 10th, 2003. The kernel function used here is the smooth cut-off biexponential function given in Eq.~(\ref{Eq:biexponential:kernel}). We rescale the instantaneous intensity in every minute. In order to observe the goodness of fitting, we also chart the real intensity, which is the number of aggressive market buy orders in every minute. It shows that our model describes the intensity dynamics quite well.

\begin{figure}
\centering
  \includegraphics[width=0.25\linewidth]{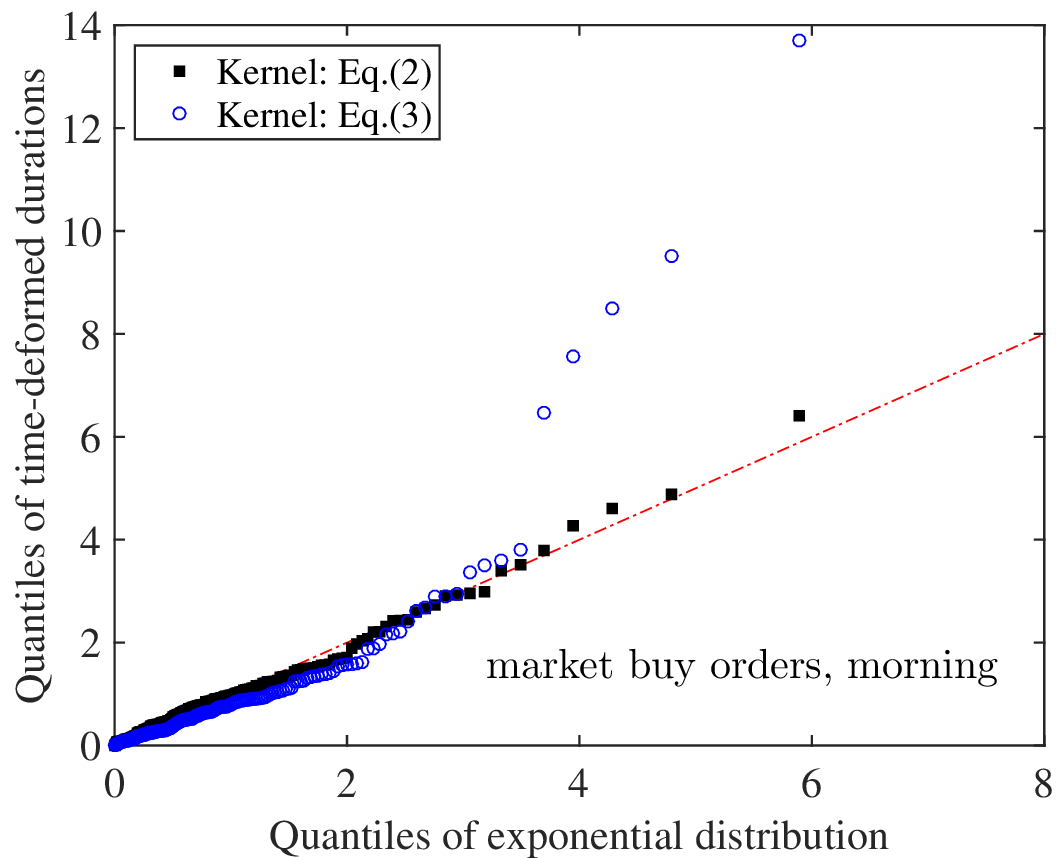}
  \includegraphics[width=0.25\linewidth]{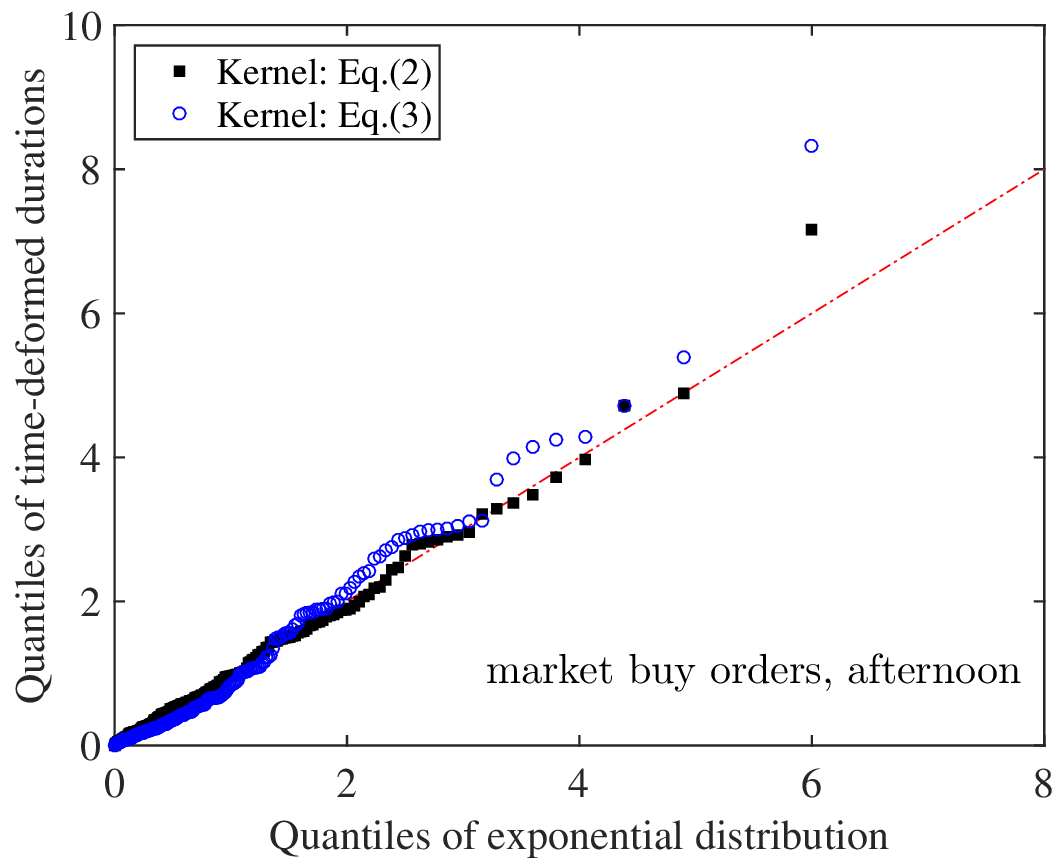}\\
  \includegraphics[width=0.25\linewidth]{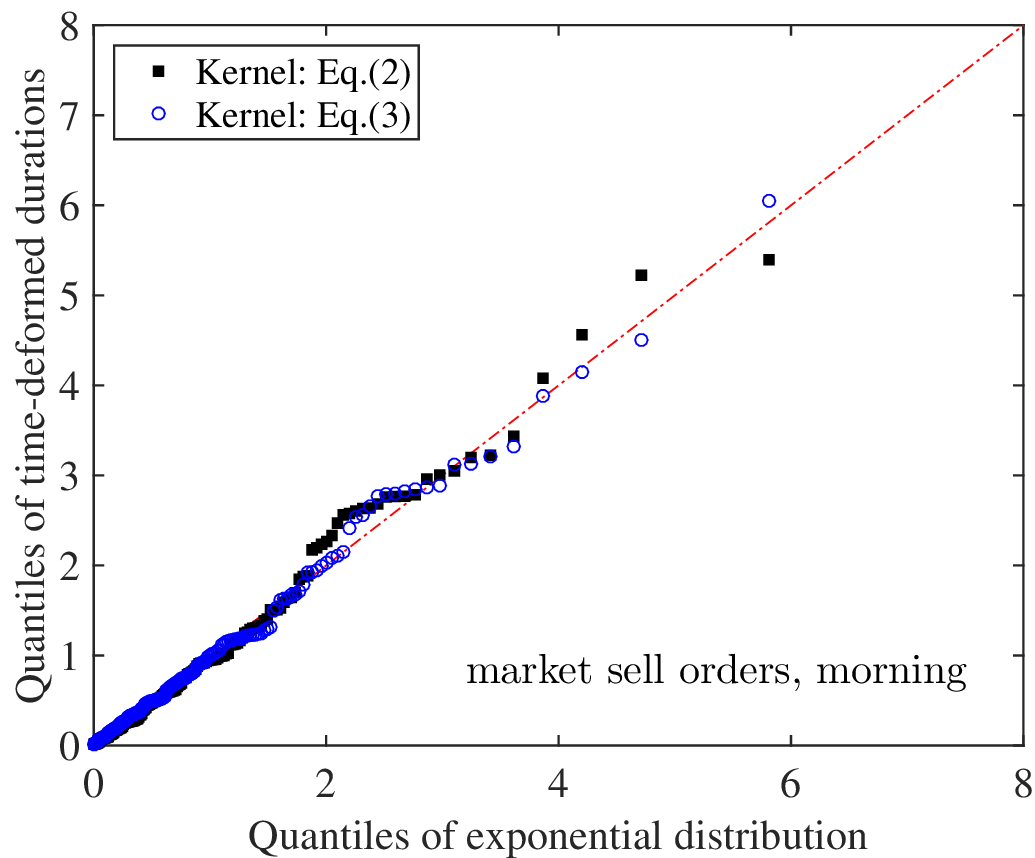}
  \includegraphics[width=0.25\linewidth]{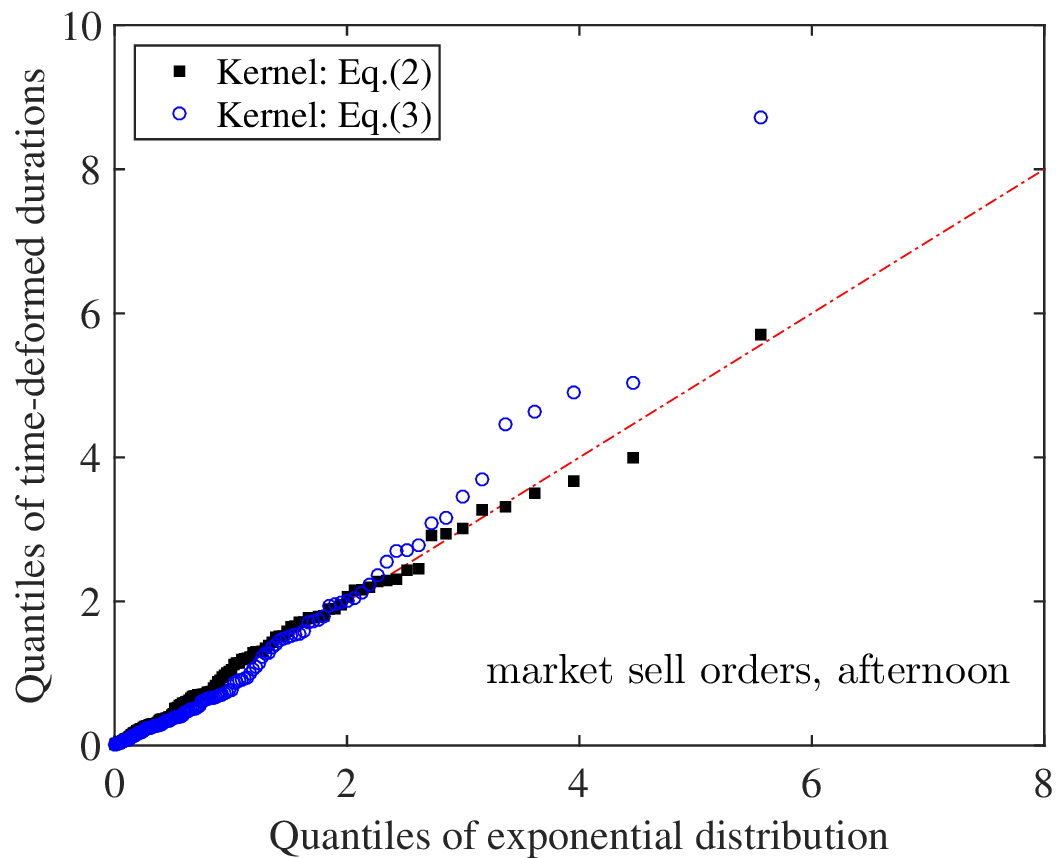}
  \caption{(color online) QQ plots (April 10th, 2003) of time-deformed durations, i.e. residuals, against an exponential distribution of parameter 1 for the two kernel functions in Eq.~(\ref{Eq:biexponential:kernel}) and Eq.~(\ref{Eq:biexponential:kernel:sum}) Upper left: aggressive market buy orders in the morning. Upper right: aggressive market buy orders in the afternoon. Lower left: aggressive market sell orders in the morning. Lower right: aggressive market sell orders in the afternoon.}
  \label{Fig:QQ}
\end{figure}

The QQ plots of the time-deformed durations defined in Eq.~(\ref{Eq:mis:specification}) on April 10th, 2003 are presented in Fig.~\ref{Fig:QQ}. We carry out the tests on both two types of market orders in either the morning or afternoon sessions. It is found that all the point collapse to the corresponding diagonals, indicating the Gaussianity of the data. Therefore, all the four fits are rather satisfactory against the theoretical exponential quantiles. This suggests our Hawkes model with the kernel in Eq.~(\ref{Eq:biexponential:kernel}) describes the data correctly.
For comparison, we also present in Fig.~\ref{Fig:QQ} the QQ plots of time-deformed durations when the kernel in Eq.~(\ref{Eq:biexponential:kernel:sum}) is used. We find that, except for the case of market sell orders in the morning, the time-deformed durations are obviously not consistent with the exponential distribution. As for the case of market sell orders in the morning, it shows a good fit due to the fact that the estimated second term in the kernel is too small. More specifically, we obtain that $\alpha_{22}=0.0089$ and $\beta_{22}=0.1733$. If the time has passed 20 seconds since the last market sell order arrival ($u$=20), the first term is $e^{-\alpha_{22}u}=0.8369$ and the second term is $e^{-\beta_{22}u}=0.0312$. Therefore, the second term has little contribution to the self-excitation process, and both the sum of exponentials kernel and the subtraction of exponentials kernel provide high goodness-of-fit. However, this is \emph{not} the usual case. For the usual cases like the other three plots in Fig.~\ref{Fig:QQ}, the second term in the kernel function is essential and the kernel in Eq.~(\ref{Eq:biexponential:kernel}) gives much better goodness-of-fit than the kernel in Eq.~(\ref{Eq:biexponential:kernel:sum}). Hence, we will only consider the smooth cut-off kernel in the following analyses.

\begin{figure}
\centering
  \includegraphics[width=0.45\linewidth]{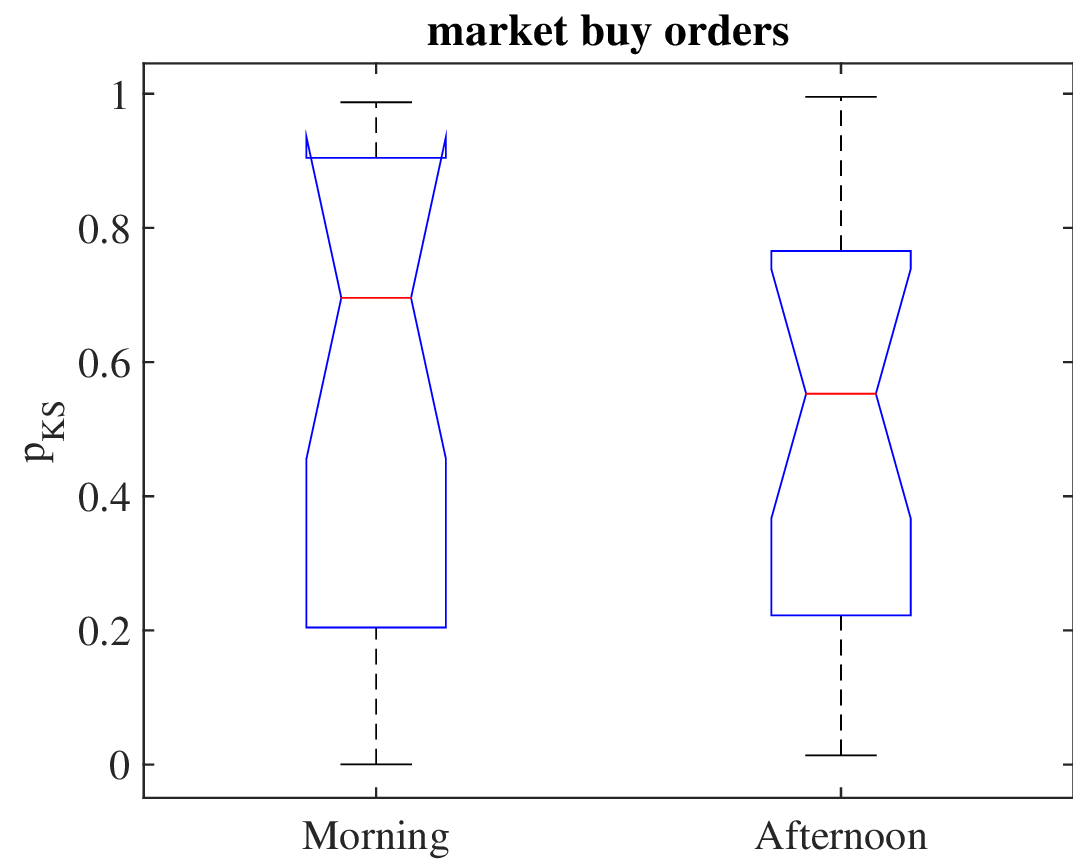}
  \includegraphics[width=0.45\linewidth]{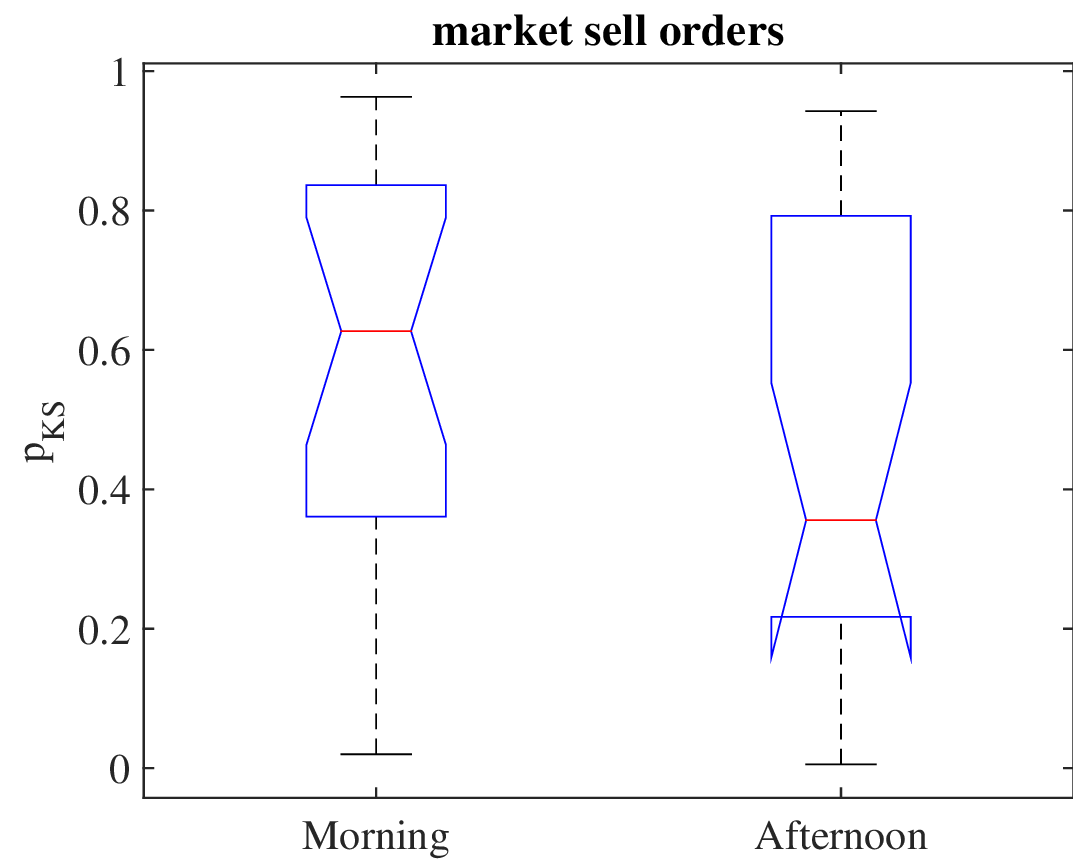}
  \caption{(color online) The $p$-values of the Kolmogorov-Smirnov test on the 21 sample days. Left: aggressive market buy orders. Right: aggressive market sell orders.}
  \label{Fig:KS:Test}
\end{figure}

Now we use the Kolmogorov-Smirnov test to analyze the goodness of fit for all sample days. Fig.~\ref{Fig:KS:Test} shows the box plot of the $p$-values of Kolmogorov-Smirnov test on all 21 sample days. This demonstrates that, with rare exceptions, almost of the samples pass the Kolmogorov-Smirnov test by a large margin. This further confirms that our bivariate Hawkes model with smooth cut-off kernels fits the market order events correctly.

\begin{figure}[h]
\centering
  \includegraphics[width=0.45\linewidth]{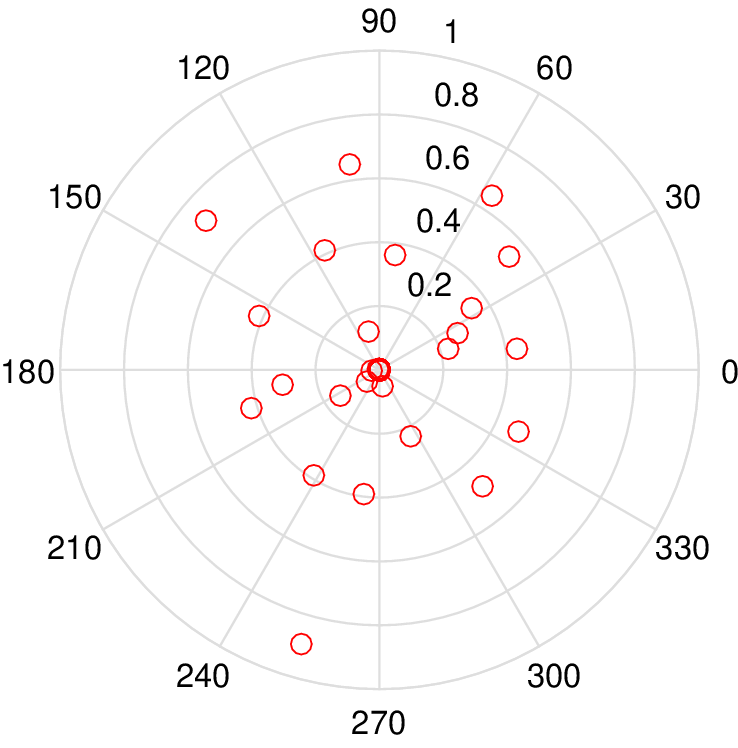}
  \caption{(color online) The spectral radiuses for 42 estimated bivariate marked Hawkes processes. It can be seen that all 42 Hawkes processes are stationary.}
  \label{Fig:spectral:radius}
\end{figure}

Then we examine whether the estimated parameters result in a stationary bivariate Hawkes process. Fig.~\ref{Fig:spectral:radius} presents the spectral radiuses for 42 estimated bivariate marked Hawkes processes, including 21 morning sessions and 21 afternoon sessions during the 21 sample days. It can be seen that all 42 spectral radiuses are strictly less than 1 and thus all 42 bivariate Hawkes processes are stationary.

\begin{figure}[ht]
\centering
  \includegraphics[width=0.45\linewidth]{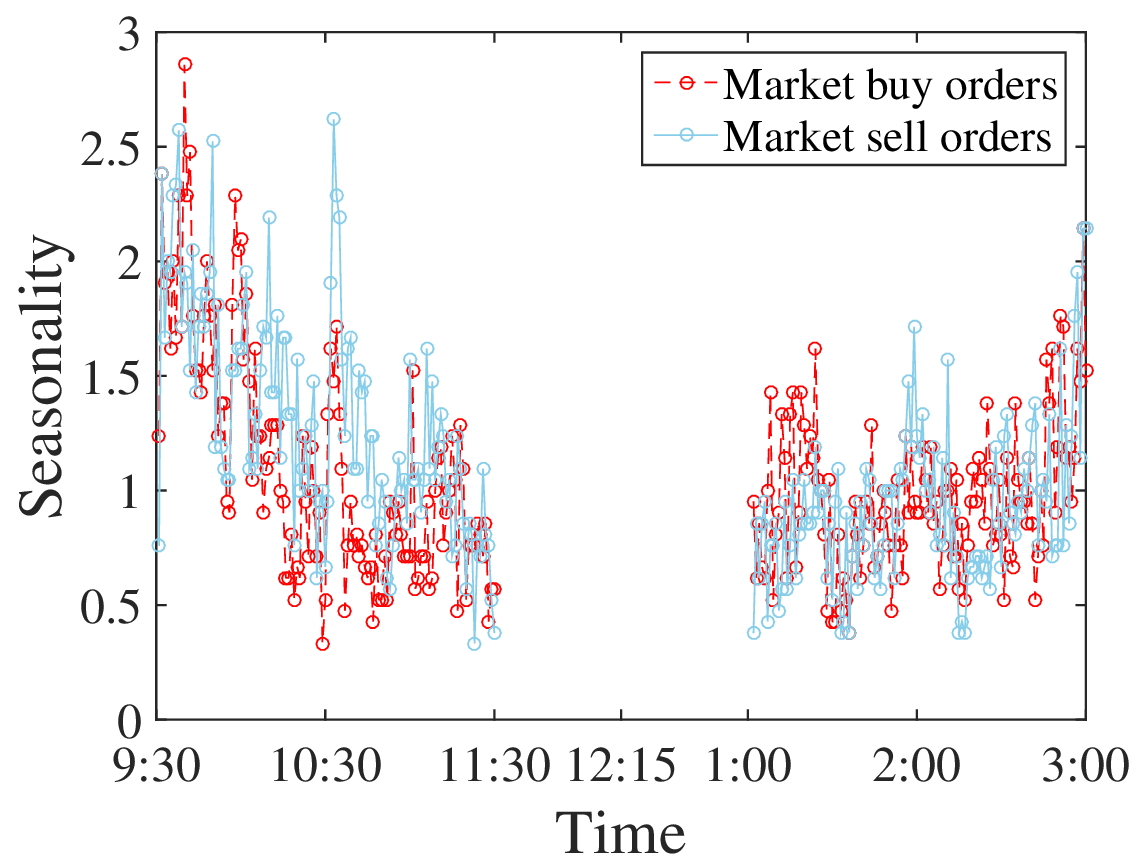}
  \includegraphics[width=0.45\linewidth,height=0.34\linewidth]{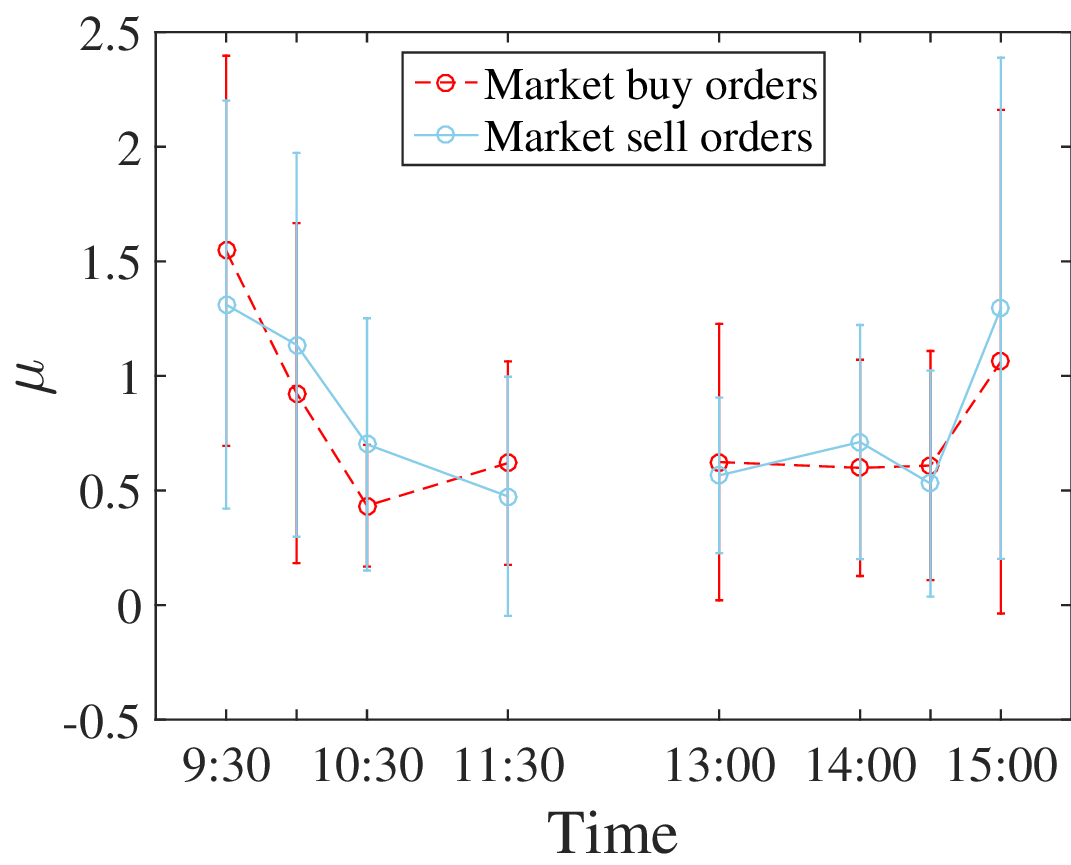}
  \caption{(color online) Average number of orders (left) and the estimated intraday baseline intensity splines (right) (events per minute). Error bars are computed for 2 standard deviations.}
  \label{Fig:season:mu}
\end{figure}

We recall that the baseline intensity $\mu(t)$ describes the arrival of exogenous events. In the left panel of Fig.~\ref{Fig:season:mu}, we first count the average market order number in every minute. The average number of orders displays the well-known $U$-shaped intraday pattern of order placement. Then, we plot the baseline intensity $\mu(t)$ in the right panel of Fig.~\ref{Fig:season:mu}. The estimated exogenous part $\mu(t)$ perfectly exhibits an intraday pattern. In addition, it is reasonable that the exogenous intensity is lower than the total intensity.

\begin{figure}
\centering
  \includegraphics[width=0.45\linewidth]{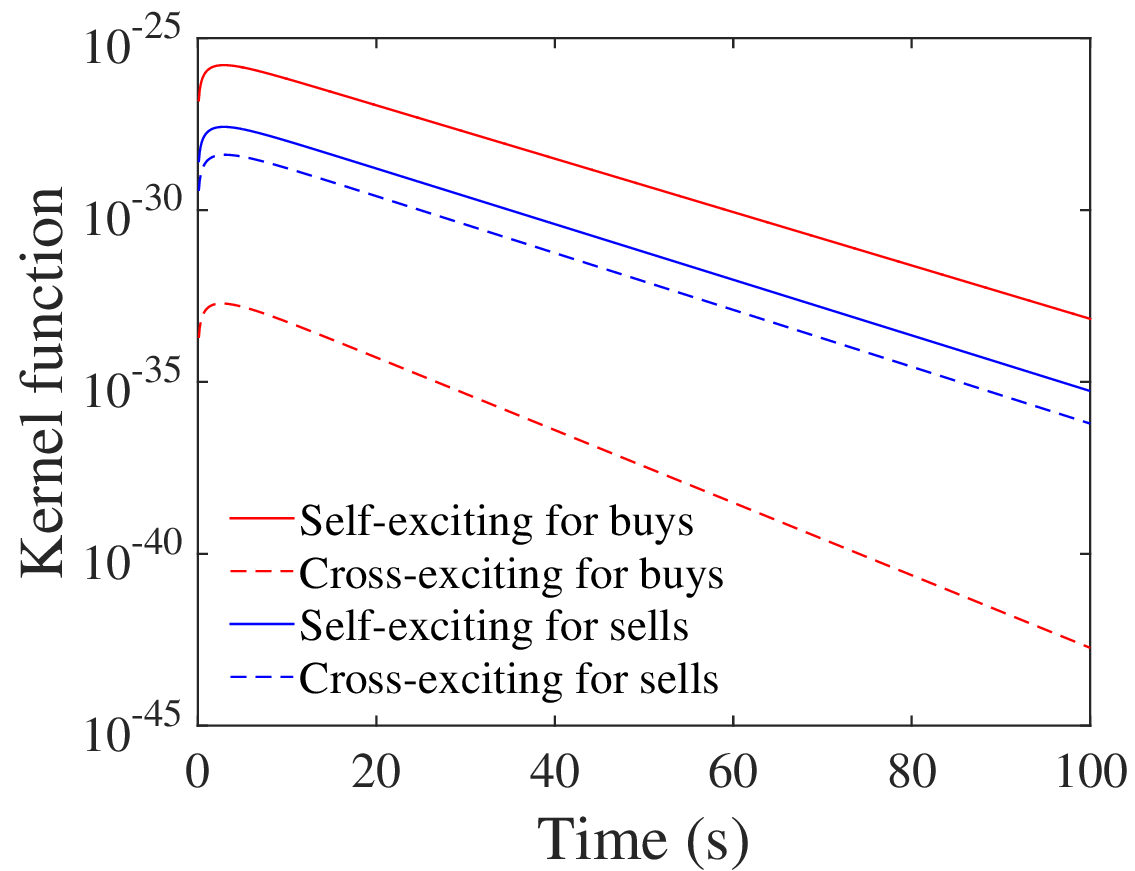}
  \caption{(color online) The estimated kernel functions $\phi_{ij}(u)$ for aggressive market buy orders and for aggressive market sell orders. The coefficients used in the kernel functions are the average values of 42 estimations.}
  \label{Fig:kernel:function}
\end{figure}

The endogenous intensity depends on self- and cross-exciting kernel functions. In Fig.~\ref{Fig:kernel:function}, we show the four estimated kernel function $\phi_{ij}(u)$ for aggressive market buy orders and aggressive market sell orders. We fix the share volume $v=3000$ (30 lots). We find that these kernel functions have a similar pattern but the scales are remarkably different. The kernel functions representing the self-exciting impact have higher values than those representing the cross-exciting impact, especially for market buy orders. This indicates that the self-excitation plays a major role in the endogenous part of aggressive market order placement.

\section{Conclusions}
\label{Sec:Conclusion}

In this work, a bivariate marked Hawkes model is proposed to characterize aggressive market order arrivals. The order arrival intensity is marked by an exogenous part and two endogenous processes reflecting respectively the self-excitation and cross-excitation. The kernel function is crucial to characterize the endogenous self-excitation and cross-excitation. We propose and compare two types of kernel function. One is a smooth cut-off exponential function (i.e. the subtraction of two exponentials), and the other is a monotonous exponential kernel (i.e. the sum of two exponentials). We calibrate the bivariate Hawkes models with different kernel functions using order flow data of a stock traded on the Shenzhen stock exchange. The bivariate Hawkes model is well estimated when the kernel is a smooth cut-off exponential function and the parameters satisfy the stationary condition. The exogenous baseline intensity explains the $U$-shaped intraday pattern. We confirm that the order arrival intensity from the endogenous part is mainly contributed to the self-exciting process, while the cross-exciting influence is weak, especially for aggressive market buy orders.

\section*{Acknowledgement}

The paper benefited greatly from the comments of participants at the Fifth Annual Meeting of the Society for Economic Measurement, Xiamen (June 2018).
This work was supported in part by the National Natural Science Foundation of China (71501072 and 71532009) and the Fundamental Research Funds for the Central Universities (2222018218006).

\bibliographystyle{elsarticle-harv}

\end{document}